\documentclass[aps,showpacs]{revtex4}
\usepackage[T1]{fontenc}
\usepackage[latin9]{inputenc}
\usepackage{amsmath}
\usepackage{amssymb}
\usepackage{graphicx}

\makeatletter
\@ifundefined{textcolor}{}
{%
 \definecolor{BLACK}{gray}{0}
 \definecolor{WHITE}{gray}{1}
 \definecolor{RED}{rgb}{1,0,0}
 \definecolor{GREEN}{rgb}{0,1,0}
 \definecolor{BLUE}{rgb}{0,0,1}
 \definecolor{CYAN}{cmyk}{1,0,0,0}
 \definecolor{MAGENTA}{cmyk}{0,1,0,0}
 \definecolor{YELLOW}{cmyk}{0,0,1,0}
}

\makeatother

\begin{document}

\preprint{This line only printed with preprint option}

\title{New Approaches in designing a Zeeman-Slower}

\author{Ben Ohayon}

\email{ben.ohayon@mail.huji.ac.il}

\affiliation{Racah Institute of physics, Hebrew University, Jerusalem 91904, Israel}

\author{Guy Ron}

\email{gron@racah.huji.ac.il}

\affiliation{Racah Institute of physics, Hebrew University, Jerusalem 91904, Israel}
\begin{abstract}
We present two new approaches for the design
of a Zeeman-Slower, which rely on optimal compliance with the adiabatic following condition and are applicable to a 
wide variety of systems. The first approach is an analytical one, based on the assumption that the noise in the system is position independent. 
When compared with the traditional approach, which requires constant deceleration, for a typical system, we show an improvement of $\sim$ 10\% in the maximal 
capture velocity, allowing for a larger slower acceptance, or a reduction of $sim$ 25\% in slower length, allowing for a simpler design and a better collimated beam.
The second approach relies on an optimization of a system in which the magnetic field and the noise profile are well known. As an example, we use our 12-coil modular 
design and show an improvement of $sim$ 9\% in maximal capture velocity or, alternatively, a reduction of $\sim$ 33\% in slower length as compared with the traditional approach.
\end{abstract}


\maketitle

\section{introduction}

Systems utilizing atomic beams frequently require those beams to be decelerated,
for example, in order to reduce the beam velocity to below the capture velocity
of a magneto optical traps. Several schemes for slowing down neutral atomic beams 
are known, these include mechanical slowing~\cite{doi:10.1021/jp002640u}, collisions with cold
background gas~\cite{PhysRevA.52.R2515}, pulsed laser fields~\cite{PhysRevLett.93.243004}, 
pulsed electric fields~\cite{PhysRevLett.83.1558},
pulsed magnetic fields~\cite{Narevicius:2008zz}, and Zeeman deceleration~\cite{PhysRevLett.48.596},
the subject of this communication. The Zeeman slower operates
by matching a spatially varying magnetic field to compensate for the change 
in the doppler shift of the decelerated beam, permitting the use of a single frequency 
laser.

The profile of this magnetic field is usually selected such that
the atoms undergo constant deceleration. A design for an optimal
coil shape which will produce this field has been presented~\cite{Dedman20045136},
but to the best of our knowledge, there is no published information examining
the optimum field shape. Even though the constant-deceleration approach
is mathematically simple, we show that it is far from ideal, and that other 
approaches are capable of optimizing various parameters of the experimental 
system. We present here two new approaches, both of which rely on satisfying
the 'Adiabatic Following'~\cite{Napolitano1990110} 
condition in the best possible manner.

The first approach is an analytical one, similar to the commonly used
constant deceleration approach and is useful for the design of systems
where the noise is independent of position along the slower.

The second approach allows to optimize an already designed 
system where the magnetic field profile and noise are well known. 
As an example of such a system we consider a design used in our lab
which aims to slow a metastable Neon beam for the purpose of trapping
it in a magneto-optical-trap.

\section{Background}

The dissipative force exerted by classical light with wave-number
$k$, on a two level system with line-width $\Gamma$, in a steady
state is 
\begin{equation}
F=\mbox{\ensuremath{\frac{\hbar k\Gamma}{2}\frac{s_{0}}{1+s_{0}+\left(4\delta/\Gamma\right)^{2}}}},\label{eq:force}
\end{equation}
where $s_0=I_l/I_S$ is the saturation parameter, $I_l$ is the laser intensity,   
$I_{s}=\hbar c\Gamma k^{3}/12\pi$ is the saturation intensity, and 

\begin{equation}
\delta\left(v,z\right)=\delta_{0}-kv+\frac{\mu'}{\hbar}B\left(z\right),\label{eq:det}
\end{equation}
is the doppler and Zeeman shifted laser detuning,
where $\delta_{0}$ is the difference between the laser and the atomic
transition frequency and $\mu'$ is the magnetic moment of the transition.

By requiring the detuned laser to be close to 
atomic resonance, $\delta\simeq0$
in (\ref{eq:det}) we arrive at an approximate relation between velocity
and magnetic field,

\begin{equation}
v\left(z\right)=\frac{\delta_{0}}{k}+\frac{\mu'}{\hbar k}B\left(z\right)\label{eq:profile}.
\end{equation}

In order to slow the atomic beam from a initial velocity $v_{i}$ to some final
velocity $v_{f}$, taking an on resonance laser, $\delta_{0}=0$,
we immediately obtain the extrema of the required magnetic field,
\begin{equation}
B_{max}=B\left(z=0\right)=\frac{\hbar k}{\mu'}v_{i};\, B_{min}=B\left(z=z_{max}\right)=\frac{\hbar k}{\mu'}v_{f}\label{eq:extrema}.
\end{equation}

Introducing an overall detuning of the laser $\delta_{0}\neq0$ is
compensated by an overall shift of the field, this is used for example
when designed a 'midfield zero' (also called a 'spin flip') slower which
has some advantageous features~\cite{tempelaars2001trapping}.
Nevertheless, from hereon we assume $\delta_{0}=0$, the general result 
is easily derived by the substitution $B\rightarrow B-\frac{\hbar}{\mu'}\delta_{0}$.

Eq. (\ref{eq:extrema}) related the magnetic field extrema 
to two important parameters, the maximal velocity, $v_{i}$, and the
length of the slower, $z_{max}$.

It is apparent that $v_{i}$ should be large in order to slow a
large fraction of the initial velocity distribution of the atomic beam, 
the need for a short slower is a somewhat more subtle. When interacting
with the laser the atoms experience
a random walk in velocity space resulting from the spontaneous emission
of photons from the laser. This results in a non zero RMS transverse
velocity or 'transverse heating' that contributes to decollimation
of the beam. We are thus driven to make the
slower as short as possible.

\section{Adiabatic Following}

Eq. (\ref{eq:profile}) relates the deceleration to the field gradient, however,
Eq. (\ref{eq:force}) enforces a strict
limit on the maximum deceleration,
\begin{equation}
a_{max}=\frac{\hbar k\Gamma}{2m}\frac{s_{0}}{1+s_{0}}\label{eq:amax}.
\end{equation}

By imposing the condition, $a=-\frac{\partial v}{\partial z}v\leq a_{max}$
and substituting velocity with the field using (\ref{eq:profile})
we obtain a constraint on the field,

\begin{equation}
-\frac{\mu'}{\hbar k}\frac{\partial B}{\partial z}\leq\frac{a_{max}}{\frac{\mu'}{\hbar k}B}\label{eq:ad}.
\end{equation}

Equation (\ref{eq:ad}) is the 'adiabatic following' condition, see
\cite{Napolitano1990110} for a more thorough discussion.

The magnetic field profile used should span from $B_{max}$ to $B_{min}$ while maintaining
the condition (\ref{eq:ad}). It is tempting to equate the two sides of (\ref{eq:ad}), and solve to get,
\begin{equation}
B\left(z\right)=\frac{\hbar k}{\mu'}\sqrt{v_{i}^{2}-2a_{max}\cdot z}\label{eq:sqrt}.
\end{equation}

This solution is also derived by demanding a constant and maximal
deceleration. The slower length is found by the minimal field from
(\ref{eq:extrema}) to be $z_{max}=\frac{v_{i}^{2}-v_{f}^{2}}{2a_{max}}$. 

While this allows a short slower, any fluctuation from this field will result in violation 
of (\ref{eq:ad}) and loss of the atoms from
the slowing process. In order to take into account such fluctuations we introduce a positive
'noise parameter', $\alpha$, which parametrizes the stability of the system
under such fluctuations,
\begin{equation}
\alpha\left(z\right)=\frac{a_{max}}{\frac{\mu'}{\hbar k}B}+\frac{\mu'}{\hbar k}\frac{\partial B}{\partial z}\label{eq:alpha}.
\end{equation}

The noise in a system can come from fluctuations in laser intensity,
see (\ref{eq:amax}), stray magnetic fields, fluctuations in currents
producing the field and especially the ability to make a desired magnetic
field in the lab.

A common approach for allowing noise in the system is to stretch
the field (\ref{eq:sqrt}) by taking $a=\eta\cdot a_{max}$, where
$\eta$ is a parameter (usually $0.5\leq\eta\leq0.7$), termed
the 'Design Parameter'. This increases the length of the required 
field, $z_{max}=\frac{v_{i}^{2}-v_{f}^{2}}{2\eta a_{max}}$. 

Inserting in (\ref{eq:alpha}) yields,
\begin{equation}
\alpha\left(z\right)=\left(1-\eta\right)\frac{a_{max}}{\sqrt{v_{i}^{2}-2\eta a_{max}z}}\label{eq:alpha2}.
\end{equation}

For $\eta=1$, we have $\alpha=0$, not allowing for any noise in the system.
For $\eta<1$; at the slower entrance, the allowed noise is $\left(1-\eta\right)a_{max}/v_{i}$,
increasing monotonously with position along the
slower. In most systems, and up to a good approximation, the noise is independent
of position. Thus, from (\ref{eq:alpha2}), the constant deceleration approach is
not ideal.

We now introduce two alternative approaches and show that they outperform 
again the traditional constant deceleration approach.

\section{Analytical Approach}
In this section we present a new approach for designing an 
optimal field profile for a system with position
independent noise.

For $\alpha\left(z\right)=\alpha$, the general solution to equation
(\ref{eq:alpha}) reads,
\begin{equation}
B\left(z\right)=\frac{\hbar k}{\mu'}\frac{a_{max}}{\alpha}\left(W\left[\left(\frac{v_{i}\alpha}{a_{max}}-1\right)e^{\frac{z\alpha^{2}+v_{i}\alpha}{a_{max}}-1}\right]+1\right)\label{eq:lambert},
\end{equation}

where $W$ is the Lambert W function~\cite{LambertW} defined as the function which solves
$z=W\left(z\right)e^{W\left(z\right)}$.

For $\alpha\rightarrow0$ it is easy to check using a series expansion
near $-e^{-1}$,
\begin{equation}
W\left(z\right)\simeq-1+\sqrt{2\left(e\times z+1\right)},
\end{equation}
that, 
\begin{equation}
B\left(z\right)\rightarrow\frac{\hbar k}{\mu'}\sqrt{v_{i}^{2}-2za_{max}},
\end{equation}
which fits (\ref{eq:sqrt}).

As with the derivation of the traditional approach, we insert (\ref{eq:extrema}) in
(\ref{eq:lambert}) to obtain the relationship between length and maximal
capture velocity,
\begin{eqnarray}
&&z_{max}=\frac{a_{max}}{\alpha^{2}}\log\left(\frac{1-\frac{v_{f}\alpha}{a_{max}}}{1-\frac{v_{i}\alpha}{a_{max}}}\right)-\frac{v_{i}-v_{f}}{\alpha}\label{eq:lamlen},\\
&&v_{i}=\frac{a_{max}}{\alpha}\left(W\left[\left(\frac{v_{f}\alpha}{a_{max}}-1\right)exp\left(-\frac{z_{max}\alpha^{2}}{a_{max}}+\frac{v_{f}\alpha}{a_{max}}-1\right)\right]+1\right)\label{eq:lamvel}
\end{eqnarray}

In order to elucidate the advantages of such an approach let us consider a
typical example. We consider the case of a metastable neon beam,
interacting with a laser of intensity $s_{0}=3$. The optical
transition of interest, $^{3}P_{2}\rightarrow{}^{3}D_{3}$, has $k=9.8\cdot10^{6}m^{-1}$,
and $\Gamma=8.2\left(2\pi\right)Mhz$. From (\ref{eq:amax}), $a_{max}=6\cdot10^{5}\, m/s^{2}$.
The magnetic moment of the transition is $\mu'=\mu_{B}$, where $\mu_B$ is the Bohr magneton.

For a desired slower length length of $z_{max}=1\, m$, a final velocity of, $v_{f}=0$, and in the 
presence of noise
that allows a design parameter of $\eta=0.5$, the maximal initial velocity
one can slow using the conventional approach is $v_{i}=\sqrt{2z_{0}\eta a_{max}}=775\, m/s$.

The minimum noise parameter is, $\alpha=\left(1-\eta\right)\frac{a_{max}}{v_{i}}=387\mbox{\ensuremath{s^{-1}}}$.
Substituting into (\ref{eq:lamvel}) we obtain the
maximum velocity allowed with the analytical approach, $v_{i}=854\, m/s$,
significantly larger.

On the other hand, for a desired initial velocity of $v_{i}=775\, m/s$,
and $\eta=0.5$, the required length obtained from (\ref{eq:lamlen}) is $z_{max}=0.77\, m$.
Allowing a reduction of $\sim$25\% is the slower length.

\begin{figure}[ht]
\includegraphics[width=0.75\textwidth]{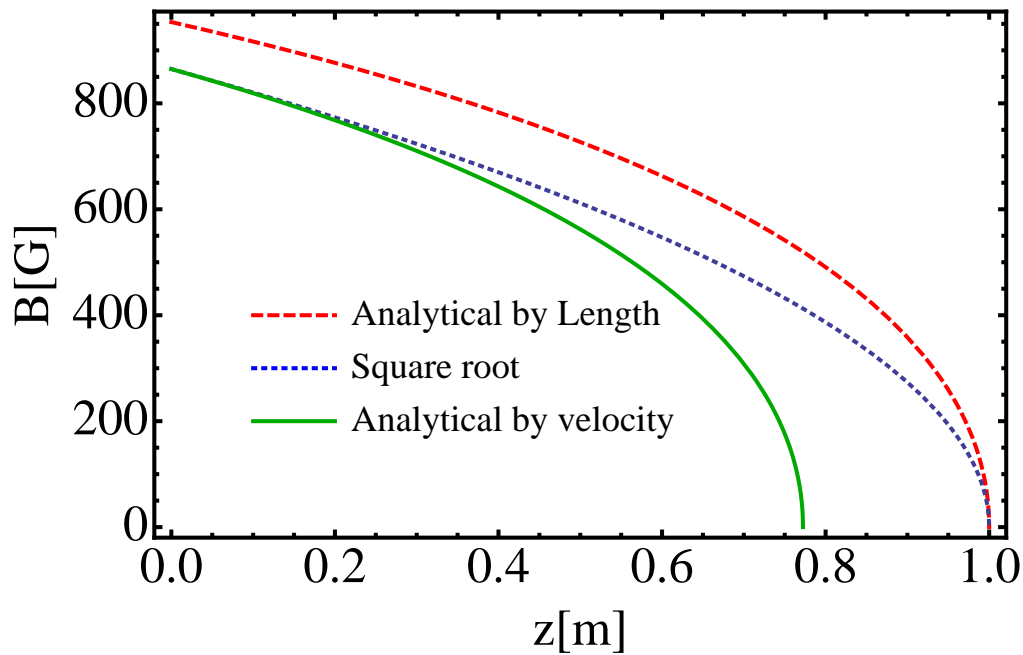}

\caption{\label{analyt}
(color online) A comparison of the three calculated Zeeman slower fields. The constant acceleration (square root) 
field (blue, dotted), the Lambert field designed for a known slower length (red, dashed), and the 
Lambert field designed for a specified capture velocity (green, solid).}
\end{figure}

It is clear that the analytical approach yields better results, both
in maximal capture velocity and in length. A Zeeman slower
designed to have this field profile can be tailored for a higher capture velocity,
less decollimation, or a combination of both. Fig. \ref{analyt} shows the three different field profiles
described in the text.

\section{System Optimization Approach}

In the previous section we described an approach to 
slower design which allows an optimal field profile 
for a position independent noise profile.This fits most of the 'tapered
solenoid' systems, where it is quite easy to make a desired target
field and the noise is mostly from fluctuations in the current.

In recent years, several systems have been designed, including
our planned apparatus~\cite{OurZeeman}, which
have more degrees of freedom on the one hand, but are harder to fit
to a target field on the other. We now show how to optimize
a known system, using our own as an example.

As a first step we write the magnetic field as a function of the
controllable system parameters (currents, winding numbers, step-motor
position (see \cite{Reinaudi:12}) etc.),
\begin{equation}
B=B\left(\vec{\theta},z\right)\label{eq:field-constrol}
\end{equation}

Now some magnetic field measurements should be done in order to determine
how well this calculated (or simulated) field represents the real
field in the lab. Any inconsistency between the two should be considered
as a part of the 'noise'.

From (\ref{eq:alpha}), assuming $\alpha$
is position dependent, we can quantify a 'noise parameter'
in units of $G/m$ as a function of the controllable parameters,
\begin{equation}
Y\left(\vec{\theta}\right)=\frac{\hbar k}{\mu'}\min\limits _{0\leq z\leq z_{max}}\alpha\left(B\left(\vec{\theta},z\right)\right)
\end{equation}

We use the minimum noise parameter to avoid the atoms falling out of the adiabatic
following condition, Eq. (\ref{eq:ad}), at any point along the slowing process.

As a next step we measure or simulate fluctuations in the controllable parameters, in order
to determine how they affect the 'noise parameter,
and select the smallest, $Y_{0}$, that takes them into account.
This will be our optimization constraint,
\begin{equation}
Y\left(\vec{\theta}\right)\geq Y_{0}\label{eq:const}.
\end{equation}
The yield function will be the parameter we wish to maximize.

Let us consider a specific system. Our group is building
a midfield-zero modular Zeeman slower, consisting 
of several independently controlled, identical coils. 
We select a detuning of, $\delta_{0}=400\left(2\pi\right)Mhz$,
which from (\ref{eq:profile}) determines the velocity in the
zero crossing point to be, $v_{c}=\delta_{0}/k=256\, m/s$. 
Following (\ref{eq:field-constrol}) we write,
\begin{equation}
B\left(\vec{I},z\right)=\underset{n}{\sum}I_{n}\times f\left(z-d\right)\label{eq:realf},
\end{equation}
where $I_{n}$ is the current in the $n^{th}$ coil, and $d=77\, mm$ is the
distance between coil centers. The function $f$ is an analytical
fit to the field of a single coil at a current of 1 A. Comparing this calculated
field to the one we measure at the lab gives good agreement up to
$\sim0.5\%$. To be on the safe side we allow about 1\% noise
in the system. Varying the currents by up to 1\% corresponds in our
system to approximately $Y_{0}=1\, G/cm$

Once we have determined the constraint  we can maximize the capture velocity.
Our  yield function is, 
\begin{equation}
v_{i}\left(\vec{I}\right)=\frac{\mu'}{\hbar k}\max\limits _{0\leq z\leq z_{max}}B\left(\vec{I},z\right)+\frac{\delta_{0}}{k}\label{eq:maxv}.
\end{equation}
We are thus fitting for the maximum of a constrained (eq (\ref{eq:const}))
nonlinear multivariable function (\ref{eq:maxv}). We use Matlab's
'fmincon' built in function which uses the 'active set' algorithm
in a similar manner to the one described by \cite{MR634376}.

\begin{figure}[ht]
\includegraphics[width=0.75\textwidth]{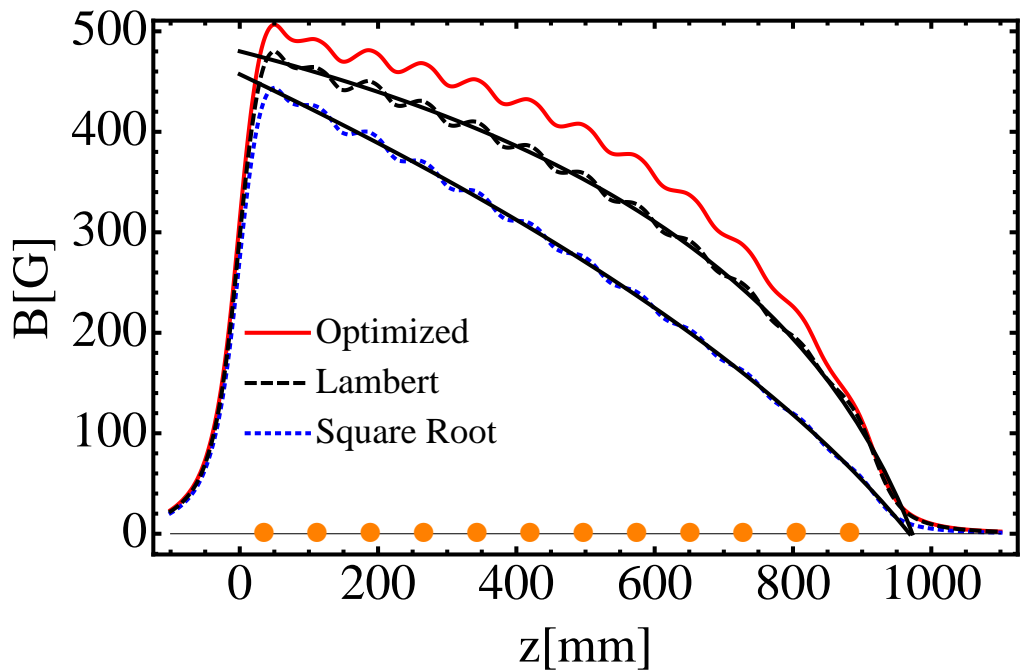}

\caption{\label{Opt12}
(color online) A comparison of the three fitted Zeeman slower fields for a constant 
length slower. The constant acceleration (square root) 
field (blue, dotted), the Lambert field designed for a known slower length (black, dashed), and the 
optimal field (red, solid). The coil center positions are also shown (orange, dots)}
\end{figure}

Figure \ref{Opt12} shows the result of optimizing for a fixed slower length 
of 12 coils compared to the results of fitting to a square root target field
and to a Lambert target field, all with with the same maximal noise
$Y_{0}=1\, G/cm$. The oscillations in the field profile stem from the discrete 
nature of the system and are taken into account implicitly in the noise parameter. 
In all cases the adiabatic following condition is satisfied.  
The optimized maximum field is $506\, G$, corresponding 
to a capture velocity of $v_{i}=710\, m/s$, whereas the Lambert field
allows $v_{i}=682\, m/s$, and the traditional (square root) field only 
permits $v_{i}=654\, m/s$.

\begin{figure}[ht]
\includegraphics[width=0.75\textwidth]{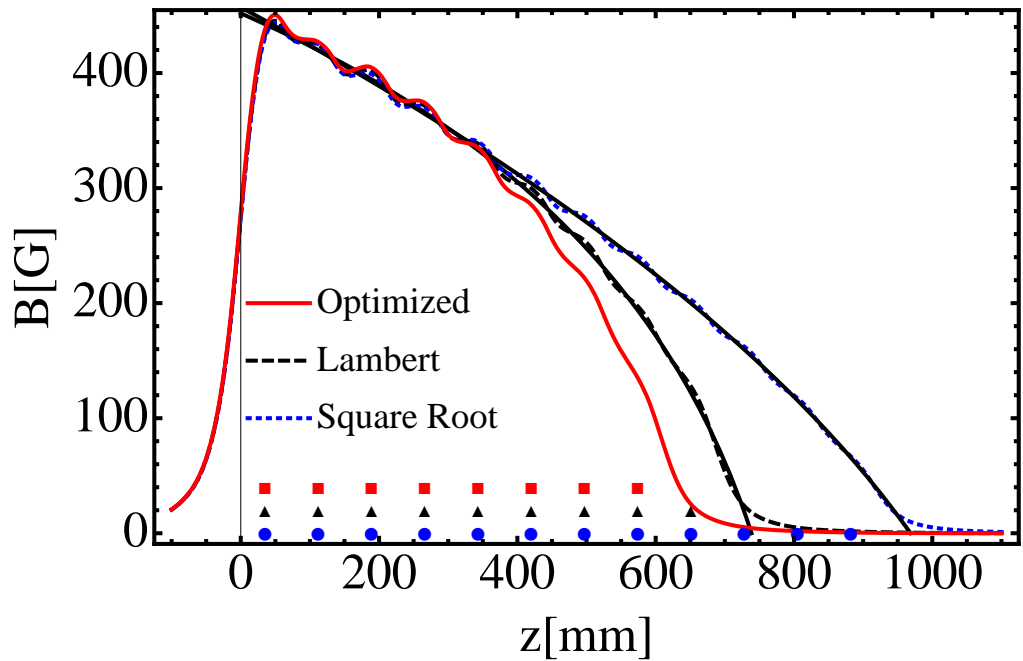}

\caption{\label{LessCoils}
(color online) A comparison of the three fitted Zeeman slower fields for a fixed 
capture velocity. The constant acceleration (square root) 
field (blue, dotted), the Lambert field (black, dashed), and the 
optimal field (red, solid). The coil center positions are also shown for each of the 
three fields (circles, triangles, and squares, respectively)}
\end{figure}

We may also use this method to design a shorter slower. We take for 
$v_{i}=654\, m/s$, which is the maximal velocity for a 12-coil slower designed with the
traditional field approach to withstand noise of $Y_{0}=1\, G/cm$.
At this initial velocity we can use the Lambert target field with 9 coils,
while still maintaining $Y_{0}\geq1\, G/cm$. When we optimize the system
under the constraint,
\begin{equation}
v_{i}\leq\frac{\mu'}{\hbar k}\max\limits _{0\leq z\leq z_{max}}B\left(\vec{I},z\right)+\frac{\delta_{0}}{k},
\end{equation}
and the yield function $Y\left(\vec{I}\right)$, we arrive at a slower
with 8 coils. The results, along with the coil center positions, are shown
in Fig. \ref{LessCoils}.

\section{summary}
In summary we have introduced two new approaches for the design of a Zeeman slower
which rely on optimal compliance with the adiabatic following condition. These approaches
allow to optimize system parameters, such as the capture velocity and the slower length.

We have shown that these two approaches yield better results than
the traditionally used square-root (constant deceleration) approach.
We note that the two approaches are complementary in that
the first is useful for the design a system and the second one
is useful for optimizing it after some measurements, such as the magnetic
field and the noise, have been taken.

\section{Acknowledgements}
This work was supported by the Israeli Science Foundation under ISF grant 177/11.

\bibliographystyle{apsrev}

\end{document}